\documentclass{article}

\begin{document}

\title{\bf \Large Classification of Static Plane Symmetric Spacetimes
according to their Matter Collineations}

\author{M. Sharif \thanks{Present Address: Department of Mathematical Sciences,
University of Aberdeen, Kings College, Aberdeen AB24 3UE Scotland,
UK. $<$msharif@maths.abdn.ac.uk$>$}
\\ Department of Mathematics, University of the Punjab,\\ Quaid-e-Azam Campus
Lahore-54590, PAKISTAN.\\ $<$hasharif@yahoo.com$>$}

\date{}

\maketitle

\begin{abstract}
In this paper we classify static plane symmetric spacetimes
according to their matter collineations. These have been studied
for both cases when the energy-momentum tensor is non-degenerate
and also when it is degenerate. It turns out that the
non-degenerate case yields either {\it four}, {\it five}, {\it
six}, {\it seven} or {\it ten} independent matter collineations in
which {\it four} are isometries and the rest are proper. There
exists three interesting cases where the energy-momentum tensor is
degenerate but the group of matter collineations is
finite-dimensional. The matter collineations in these cases are
either {\it four}, {\it six} or {\it ten}.
\end{abstract}


\date{}

\newpage

\section{Introduction}

There exists a large body of literature on classification of
spacetimes according to their isometries or Killing vectors (KVs)
and the groups admitted by them [1]-[4]. These investigations of
symmetries played an important role in the classification of
spacetimes, giving rise to many interesting results with useful
applications. As curvature and Ricci tensors play a significant
role in understanding the geometric structure of metrics, the
energy-momentum tensor enables us to understand the physical
structure of spacetimes. Symmetries of the energy-momentum tensor
(also called matter collineations) provide conservation laws on
matter fields. These enable us to know how the physical fields,
occupying in certain region of spacetimes, reflect the symmetries
of the metric [5].

Some recent literature [6]-[12] shows keen interest in the study
of matter collineations (MCs). In one of the recent papers [12],
the study of MCs has been taken for static spherically symmetric
spacetimes and some interesting results have been obtained. In
this paper, we address the same problem for static plane symmetric
spacetimes. It turns out that static plane symmetric spacetimes
admit an MC Lie algebra of 10, 7, 6, 5 and 4 dimensions apart from
the infinite dimensional algebras.

Let $(M,g)$ be a spacetime, where $M$ is a smooth, connected,
Hausdorff four-dimensional manifold and $g$ is smooth Lorentzian
metric of signature (+ - - -) defined on $M$. The manifold $M$ and
the metric $g$ are assumed smooth ($C^{\infty}$). We shall use the
usual component notation in local charts, and a covariant
derivative with respect to the symmetric connection $\Gamma$
associated with the metric $g$ will be denoted by a semicolon and
a partial derivative by a comma. A smooth vector field ${\bf \xi}$
is said to preserve a matter symmetry [13] on $M$ if, for each
smooth local diffeomorphism $\phi_t$ associated with ${\bf \xi}$,
the tensors $T$ and $\phi^*_tT$ are equal on the domain $U$ of
$\phi_t$, i.e., $T=\phi_t^*T$. Equivalently, a vector field
$\xi^a$ is said to generate a matter collineation if it satisfies
the following equation
\begin{equation}
\pounds_{\xi}T_{ab}=0,
\end{equation}
where $\pounds$ is the Lie derivative operator, $\xi^a$ is the
symmetry or collineation vector. Every KV is an MC but the
converse is not true, in general. Collineations can be proper
(non-trivial) or improper (trivial). We define a proper MC to be
an MC which is not a KV, or a homothetic vector (HV). The MC
Eq.(1) can be written in component form as
\begin{equation}
T_{ab,c} \xi^c + T_{ac} \xi^c_{,b} + T_{cb} \xi^c_{,a}=
0,~~(a,b,c=0,1,2,3).
\end{equation}

A plane symmetric spacetime is a Lorentzian manifold possessing a
physical stress-energy tensor. This admits $SO(2)(\times\Re^2$ as
the minimal isometry group in such a way that the group orbits are
spacelike surfaces of constant curvature. The metric for static
plane symmetric spacetimes is given in the form [3]
\begin{equation}
ds^2=e^{\nu(x)}dt^2-dx^2-e^{\mu(x)}(dy^2+dz^2),
\end{equation}
where $\nu$ and $\mu$ are arbitrary functions of $x$. The
surviving components of the energy-momentum tensor, given in
Appendix A, are $T_{0},~T_{1},~T_{2},~T_{3}$, where $T_{3}=T_{2}$
and we have used the notation $T_{aa}=T_a$ for the sake of
simplicity.

The MC equations can be written as follows
\begin{equation}
T_{0,1}\xi^1+2T_0\xi^0_{,0}=0,
\end{equation}
\begin{equation}
T_0\xi^0_{,1}+T_1\xi^1_{,0}=0,
\end{equation}
\begin{equation}
T_0\xi^0_{,2}+T_2\xi^2_{,0}=0,
\end{equation}
\begin{equation}
T_0\xi^0_{,3}+T_2\xi^3_{,0}=0,
\end{equation}
\begin{equation}
T_{1,1}\xi^1+2T_1\xi^1_{,1}=0,
\end{equation}
\begin{equation}
T_1\xi^1_{,2}+T_2\xi^2_{,1}=0,
\end{equation}
\begin{equation}
T_1\xi^1_{,3}+T_2\xi^3_{,1}=0,
\end{equation}
\begin{equation}
T_{2,1}\xi^1+2T_2\xi^2_{,2}=0,
\end{equation}
\begin{equation}
T_2(\xi^2_{,3}+\xi^3_{,2})=0,
\end{equation}
\begin{equation}
T_{2,1}\xi^1+2T_2\xi^3_{,3} = 0.
\end{equation}
These are the first order non-linear partial differential
equations in four variables $\xi^a(x^b)$. We solve these equations
for the non-degenerate case, when
\begin{equation}
\det(T_{ab})=T_0T_1T^2_2\neq 0
\end{equation}
and for the degenerate case, where $\det(T_{ab})=0$. The rest of
the paper is organized as follows. The next section contains a
solution of MC equations when the energy-momentum tensor is
non-degenerate. In section 3, MC equations are solved for the
degenerate energy-momentum tensor and section 4 provides some
examples admitting proper MCs for the non-degenerate case.
Finally, section 5 contains a summary and discussion of the
results obtained.

\section{Matter Collineations in the Non-Degenerate Case}

In this section, we shall evaluate MCs only for those cases which
have non-degenerate energy-momentum tensor, i.e.,
$\det(T_{ab})\neq 0$. To this end, we set up the general
conditions for the solution of MC equations for the non-degenerate
case.

When we solve Eqs.(4)-(13) simultaneously, after some algebraic
computations, we arrive at the following solution
\begin{eqnarray}
\xi^0&=&-\frac{T_2}{T_0}[\frac{1}{2}(y^2+z^2)\dot{A}_1
+z\dot{A}_2+y\dot{A}_3]+A_4,\\
\xi^1&=&-\frac{T_2}{T_1}[\frac{1}{2}(y^2+z^2)A'_1+zA'_2+yA'_3]+A_5,\\
\xi^2&=&\frac{1}{2}z^2(c_1y+c_3)+z(c_2y+c_4)-\frac{1}{6}c_1y^3
-\frac{1}{2}c_3y^2+yA_1+A_3,\\
\xi^3&=&-\frac{1}{2}y^2(c_1z+c_2)-y(c_3z+c_4)+\frac{1}{6}c_1z^3
+\frac{1}{2}c_2z^2+zA_1+A_3.
\end{eqnarray}
where $c_1,c_2,c_3,c_4$ are arbitrary constants and
$A_\mu=A_\mu(t,x),~\mu=1,2,3,4,5$ are integration constants. Here
dot and prime indicate the differentiation with respect to time
and $x$ coordinate respectively. When we replace these values of
$\xi^a$ in MC Eqs.(4)-(13), we obtain the following constraints on
$A_\mu$
\begin{equation}
\frac{T'_0}{T_1}A'_i+2\ddot{A}_i=0,~~(i=1,2,3),
\end{equation}
\begin{equation}
\dot{A}_i=\sqrt{\frac{T_0}{T_2}}f_i(t),~
A'_i=\frac{\sqrt{T_1}}{T_2}g_i(t),
\end{equation}
\begin{equation}
\dot{A}_4=-\frac{T'_0}{2T_0\sqrt{T_1}}g_5(t),~
A'_4=-\frac{\sqrt{T_1}}{T_0}\dot{g}_5(t),
\end{equation}
\begin{equation}
c_1=0,~T'_2A'_1=0,
\end{equation}
\begin{equation}
\frac{T'_2}{T_2}A_5+2A_1=0,
\end{equation}
\begin{equation}
\frac{T'_2}{T_2\sqrt{T_1}}g_2(t)-2c_2=0,
\end{equation}
\begin{equation}
\frac{T'_2}{T_2\sqrt{T_1}}g_3(t)+2c_3=0,
\end{equation}
\begin{equation}
\frac{T'_2}{T_2\sqrt{T_1}}g_5(t)+2A_1=0,
\end{equation}
\begin{equation}
\frac{T'_0}{T_2\sqrt{T_1}}g_i(t)+2\ddot{A}_i=0,
\end{equation}
where $f_i(t),g_i(t),~g_5(t)$ are integration functions. Thus the
problem of working out MCs for all possibilities of $A_i,A_4,A_5$
is reduced to solving the set of Eqs.(15)-(18) subject to the
above constraints. We would solve these to classify MCs of the
plane symmetry manifolds.

From Eqs.(24)-(26), there arises two main cases: \\
\par \noindent
\par \noindent
(1) $~~~~~(\frac{T'_2}{T_2\sqrt{T_1}})'\neq 0,~~$
(2) $~~~~~(\frac{T'_2}{T_2\sqrt{T_1}})'=0$.\\
\par \noindent
\par \noindent
{\bf Case (1)}: In this case, we have $T'_2\neq 0$ and hence
Eq.(22) gives $A_1=A_1(t)$. Using these in Eq.(26), it follows
that
\begin{equation}
\frac{T'_2}{T_0\sqrt{T_1}}g_5(t)+2A_1(t)=0
\end{equation}
which implies that $g_5=0$ and $A_1=0$. Thus we have from Eqs.(21)
and (23) $A_5=0,~A_4=c_0$. Also, Eqs.(24) and (25) yield
\begin{equation}
g_2=0=g_3,~c_2=0=c_3.
\end{equation}
Now from Eqs.(19) and (20), we have
\begin{equation}
A'_j=0,~\ddot{A}_j=0,~\dot{f}_j=0,~(j=2,3)
\end{equation}
which gives
\begin{equation}
A_j(t,x)=\sqrt{\frac{T_0}{T_2}}c_jt+c_{j+2}.
\end{equation}
Since $A'_j(t,x)=0$ which implies that either\\
(a) $~~~(\frac{T_0}{T_2})'=0,~~~$ or (b)
$~~~(\frac{T_0}{T_2})'\neq 0$.

In the first case 1(a), we have the following MCs
\begin{eqnarray}
\xi_{(1)}=\partial_t,~~\xi_{(2)}=\partial_y,~~
\xi_{(3)}=\partial_z,~~
\xi_{(4)}=z\partial_y-y\partial_z,\nonumber\\
\xi_{(5)}=t\partial_z-\frac{T_2}{T_0}z\partial_t,~~
\xi_{(6)}=t\partial_y-\frac{T_2}{T_0}y\partial_t.
\end{eqnarray}
Thus we obtain six independent MCs in which four are the usual
isometries of the plane symmetry and the rest are the proper MCs.
The MCs for the case 1(b) turns out to be the same as the minimal
isometries for the plane symmetry.
\par \noindent
\par \noindent
{\bf Case (2)}: This case implies that
$\frac{T'_2}{T_2\sqrt{T_1}}=\alpha$, where $\alpha$ is an
arbitrary constant and can have the following two
subcases according as $\alpha$ is non-zero or zero. \\
(a) $~~~~\alpha\neq 0,~~~$ (b) $~~~\alpha=0$.

For the case 2(a), we use Eqs.(20),(22),(24) and (25) so that
\begin{equation}
g_2=2c_2/\alpha,~g_3=-2c_3/\alpha,~A'_1=0,
\end{equation}
and
\begin{equation}
(\frac{T_2}{T_0})'\dot{A_i}=0.
\end{equation}
The last equation further gives us the following two possibilities:\\
(i) $~~~(\frac{T_2}{T_0})'\neq 0,~~~$ (ii)
$~~~(\frac{T_2}{T_0})'=0$.

In the first case 2a(i), Eqs.(20),(21),(23),(25),(36) and (37)
imply that
\begin{equation}
A_5=c_5,~T'_0A'_i=0,~A_5=-\frac{2c_5}{\alpha\sqrt{T_1}},~A'_4=0,
\end{equation}
and
\begin{equation}
\dot{A}_4=\frac{T'_0}{\alpha T_0\sqrt{T_1}}c_5.
\end{equation}
This last equation implies that for
$(\frac{T'_0}{T_0\sqrt{T_1}})'\neq 0$, we have the same MCs as
KVs. When $\frac{T'_0}{T_0\sqrt{T_1}}=\beta$, where $\beta$ is an
arbitrary constant, this further gives the following two subcases\\
(*) $~~~\beta\neq 0,~~~$ (**) $~~~\beta=0$.

The case 2ai(*), in addition to the usual isometries of plan
symmetry, gives the following one proper MC
\begin{equation}
\xi_{(5)}=\frac{\beta}{\alpha}t\partial_t
-\frac{2}{\alpha\sqrt{T_1}}\partial_x+y\partial_y+z\partial_z.
\end{equation}

For the case 2ai(**), we have $T_0=constant$ and we obtain the
following MCs
\begin{eqnarray}
\xi_{(5)}&=&yz+(\frac{z^2}{2}-\frac{y^2}{2}-\frac{2}{\alpha^2
T_2})\partial_z,\nonumber\\
\xi_{(6)}&=&yz-(\frac{z^2}{2}-\frac{y^2}{2}+\frac{2}{\alpha^2
T_2})\partial_y,\nonumber\\
\xi_{(7)}&=&y\partial_y+z\partial_z.
\end{eqnarray}
This implies that we have seven independent MCs in which three are
the proper MCs.

In the case 2a(ii), we obtain $T_2=\gamma T_0$, where $\gamma$ is
an arbitrary constant and this yields the following MCs
\begin{eqnarray}
\xi_{(5)}&=&\frac{1}{2}(t^2-\frac{4}{\alpha^2T_0}-\gamma
y^2-\gamma z^2)\partial_t+\frac{2}{\alpha\sqrt{T_1}}\partial_x
+ty\partial_y+tz\partial_z,\nonumber\\
\xi_{(6)}&=&\frac{1}{\gamma}tz\partial_t
+\frac{2}{\alpha^2\sqrt{T_1}}z\partial_x+yz\partial_y
-\frac{1}{2}(\frac{t^2}{\gamma}+\frac{4}{\alpha^2T_2}
+y^2-z^2)\partial_z,\nonumber\\
\xi_{(7)}&=&z\partial_t-t\partial_z,\nonumber\\
\xi_{(8)}&=&\frac{1}{\gamma}ty\partial_t
+\frac{2}{\alpha^2\sqrt{T_1}}y\partial_x
-\frac{1}{2}(\frac{t^2}{\gamma}+\frac{4}{\alpha^2T_2}
-y^2+z^2)\partial_y-yz\partial_z,\nonumber\\
\xi_{(9)}&=&y\partial_t-t\partial_y,\nonumber\\
\xi_{(10)}&=&t\partial_t+\frac{2}{\alpha\sqrt{T_1}}\partial_x
+y\partial_y+z\partial_z).
\end{eqnarray}
This shows that we have ten independent MCs including six proper
MCs.

The case 2(b) implies that $T_2=constant$ which yields that
either\\
(i) $~~~(\frac{(\sqrt{T_0})'}{\sqrt{T_1}})'=0~~~$ or (ii)
$~~~(\frac{(\sqrt{T_0})'}{\sqrt{T_1}})'\neq 0$

For the first possibility 2b(i), we have
$\frac{(\sqrt{T_0})'}{\sqrt{T_1}}=\delta$, where $\delta$ is an
arbitrary constant and gives two possibilities according as it is
non-zero or zero\\
(*) $~~~\delta\neq 0,~~~$ (**) $~~~\delta=0$.

For the case 2bi(*), we obtain the following MCs
\begin{eqnarray}
\xi_{(5)}&=&\frac{T_2}{\sqrt{T_0}}z\sin\delta t\partial_t
-\frac{T_2}{\sqrt{T_1}}z\cos\delta t\partial_x
+\frac{T_0}{\delta}\cos\delta t\partial_z,\nonumber\\
\xi_{(6)}&=&\frac{T_2}{\sqrt{T_0}}z\cos\delta t\partial_t
+\frac{T_2}{\sqrt{T_1}}z\sin\delta t\partial_x
-\frac{T_0}{\delta}\sin\delta t\partial_z,\nonumber\\
\xi_{(7)}&=&\frac{T_2}{\sqrt{T_0}}y\sin\delta t\partial_t
-\frac{T_2}{\sqrt{T_1}}y\cos\delta t\partial_x
+\frac{T_0}{\delta}\cos\delta t\partial_y,\nonumber\\
\xi_{(8)}&=&\frac{T_2}{\sqrt{T_0}}y\cos\delta t\partial_t
+\frac{T_2}{\sqrt{T_1}}y\sin\delta t\partial_x
-\frac{T_0}{\delta}\sin\delta t\partial_y,\nonumber\\
\xi_{(9)}&=&\frac{1}{\sqrt{T_0}}\sin\delta t\partial_t
-\frac{1}{\sqrt{T_1}}\cos\delta t\partial_x,\nonumber\\
\xi_{(10)}&=&\frac{1}{\sqrt{T_0}}\cos\delta t\partial_t
+\frac{1}{\sqrt{T_1}}\sin\delta t\partial_x
\end{eqnarray}
which yields ten independent MCs having six proper MCs.

In the case of 2bi(**), we have the following MCs
\begin{eqnarray}
\xi_{(5)}&=&\frac{T_2}{\sqrt{T_1}}z\partial_x
-\int{\sqrt{T_1}dx}\partial_z,\nonumber\\
\xi_{(6)}&=&\frac{T_2}{\sqrt{T_1}}y\partial_x
-t\partial_z,\nonumber\\
\xi_{(7)}&=&\frac{T_2}{\sqrt{T_1}}z\partial_t
-\int{\sqrt{T_1}dx}\partial_y,\nonumber\\
\xi_{(8)}&=&\frac{T_2}{\sqrt{T_1}}y\partial_t
-t\partial_y,\nonumber\\
\xi_{(9)}&=&\frac{1}{T_0}\int{\sqrt{T_1}dx}\partial_t
-\frac{1}{\sqrt{T_1}}t\partial_x,\nonumber\\
\xi_{(10)}&=&\frac{1}{\sqrt{T_1}}\partial_x
\end{eqnarray}
giving ten independent MCs with six proper MCs.

The case 2b(ii) further implies the following two possibilities:\\
(*)
$~~~(\frac{T_0}{2}\sqrt{T_1}(\frac{T'_0}{T_0\sqrt{T_1}})')'=0,~~~$
(**)
$~~~(\frac{T_0}{2}\sqrt{T_1}(\frac{T'_0}{T_0\sqrt{T_1}})')'\neq
0$.

For 2bii(*), we have
$\frac{T_0}{2}\sqrt{T_1}(\frac{T'_0}{T_0\sqrt{T_1}})'=\epsilon$,
where $\epsilon$ is an integration constant and gives further two
cases when\\
(+) $~~~\epsilon=0~~~$ and (++) $~~~\epsilon\neq 0$.

In the case 2bii*(+), we have $\frac{T'_0}{T_0\sqrt{T_1}}=\chi\neq
0$ and this gives the following MCs
\begin{eqnarray}
\xi_{(5)}&=&(\frac{1}{\chi T_0}-\frac{\chi}{4}t^2)\partial_t
+\frac{1}{\sqrt{T_1}}t\partial_x,\nonumber\\
\xi_{(6)}&=&(\frac{\chi}{2}t\partial_t
+\frac{1}{\sqrt{T_1}}\partial_x
\end{eqnarray}
yielding six independent MCs.

For the case 2bii*(++), we obtain
\begin{eqnarray}
\xi_{(5)}&=&(\frac{1}{T_0}\partial_t
-\frac{1}{\sqrt{T_1}}\partial_x)e^{\sqrt{\eta}t},\nonumber\\
\xi_{(6)}&=&(\frac{1}{T_0}\partial_t
+\frac{1}{\sqrt{T_1}}\partial_x)e^{-\sqrt{\eta}t}
\end{eqnarray}
giving six independent MCs.

In the case 2bii(**), we get MCs equal to the KVs.

\section{Matter Collineations in the Degenerate Case}

In this section only those cases will be considered for which the
energy-momentum tensor is degenerate, i.e., $\det(T_{ab})=0$. Thus
we would discuss the spacetimes when at least one of the $T_a$ or
their combination is zero. When $T_a=0$, we have trivially every
direction is an MC. The remaining cases can be classified as
follows:
\par \noindent
\par \noindent
1. When only one of $T_a$ is non-zero;\\
2. When two of $T_a$ are non-zero;\\
3. When three of $T_a$ are non-zero.\\
\par \noindent
\par \noindent
{\bf Case (1)}: This can further be grouped as follows:
\par \noindent
\par \noindent
(a) $~~T_0\neq 0,~~T_i=0,~~$ (b) $~~T_1\neq
0,~~T_j=0,~~(i=1,2,3),(j=0,2,3)$.

The case 1(a) yields two possibilities according as $T'_0=0$ or
$T'_0\neq 0$. For the first possibility, we get
\begin{equation}
\xi^0=c_0,~~\xi^i=\xi^i(x^a).
\end{equation}
The second possibility implies that
\begin{equation}
\xi^0=\xi^0(t),~~\xi^1=-\frac{2T_0}{T'_0}\dot{\xi}^0(t),~~
\xi^k=\xi^k(x^a),~~(k=2,3).
\end{equation}
Thus we have infinite dimensional MCs.

The case 1(b) can be solved trivially and gives
\begin{equation}
\xi^1=\frac{c_1}{\sqrt{T_1}},~~\xi^j=\xi^j(x^a)
\end{equation}
which implies infinite dimensional MCs.
\par \noindent
\par \noindent
{\bf Case (2)}: This case can be divided into the following cases:
\par \noindent
\par \noindent
(a) $~~~T_l=0,~~T_k\neq 0~~(l=0,1$ and $k=2,3)$,\\
(b) $~~~T_l\neq 0,~~T_k=0$.

In the first case, if we take $T_2=constant$, then we have the
following MCs
\begin{equation}
\xi^l=\xi^l(x^a),~~~\xi^2=c_0z+c_1,~~~\xi^3=-c_0y+c_2
\end{equation}
which gives infinite dimensional MCs. For $T'_2\neq 0$, we again
have infinite dimensional MCs given by
\begin{eqnarray}
\xi^0&=&\xi^0(x^a),\\
\xi^1&=&-\frac{T_2}{T'_2}(f'(u)+g'(u)),\\
\xi^2&=&f(u)+g(v),\\
\xi^3&=&\iota(-f(u)+g(v))+c_0,
\end{eqnarray}
where $u=y+\iota z$ and $v=y-\iota z$.

For the second case 2(b), it follows from Eqs.(4)-(7) and (9)-(10)
that $\xi^l=\xi^l(t,x),~\xi^k=\xi^k(x^a)$. Also, Eq.(8) yields
$\xi^1=\frac{f(t)}{T_1}$. If we use this value in Eqs.(4)-(5) and
eliminate $\xi^0$, we have
\begin{equation}
\ddot{f}(t)=\frac{T_0}{\sqrt{T_1}}(\frac{T'_0}{2T_0\sqrt{T_1}})'f(t).
\end{equation}
From this equation, we see that for $f=0$, we have infinite
dimensional MCs given by
\begin{equation}
\xi^0=c_0,~~~\xi^l=0,~~~\xi^k=\xi^k(x^a).
\end{equation}
For $f(t)\neq 0$, we have
\begin{equation}
\frac{\ddot{f}(t)}{f(t)}=\frac{T_0}{\sqrt{T_1}}(\frac{T'_0}
{2T_0\sqrt{T_1}})'=\alpha,
\end{equation}
where $\alpha$ is an arbitrary constant. This gives two
possibilities either $\alpha=0$ or $\alpha\neq 0$. For the first
possibility, we obtain $\frac{T'_0}{2T_0\sqrt{T_1}}=\beta$, an
arbitrary constant. This again yields the infinite dimensional MCs
given by
\begin{eqnarray}
\xi^0&=&c_1(-\frac{\beta}{2}t^2-\int{\frac{\sqrt{T_1}}{T_0}dx})
-c_2\beta t+c_0,\\
\xi^l&=&\frac{1}{T_1}(c_1t+c_2),\\
\xi^k&=&\xi^k(x^a).
\end{eqnarray}
When $\alpha\neq 0$, we have infinite dimensional MCs as follows
\begin{eqnarray}
\xi^0&=&-\frac{T'_0}{2T_0\sqrt{\alpha
T_1}}(c_1e^{\sqrt{\alpha}t}-c_2e^{-\sqrt{\alpha}t}+c_3,\\
\xi^l&=&\frac{1}{T_1}(c_1e^{\sqrt{\alpha}t}+c_2e^{-\sqrt{\alpha}t},\\
\xi^k&=&\xi^k(x^a).
\end{eqnarray}
{\bf Case (3)}: This case can be divided as follows:
\par \noindent
\par \noindent
(a) $~~~T_0=0,~~T_i\neq 0$; (b) $~~~T_1=0,~~T_j\neq 0$.

In the case 3(a), it is easy to see that Eqs.(4)-(7) imply that
$\xi^0$ is an arbitrary function of four variables while
$\xi^i=\xi^i(x,y,z)$. Further, it follows from Eqs.(8)-(11) and
(13) that
\begin{equation}
A_1(y,z)_{,kk}-(\frac{T'_2}{2T_2\sqrt{T_1}})'
\frac{T_2}{\sqrt{T_1}}A_1(y,z)=0.
\end{equation}
From here we have two possibilities either $A_1=0$ or $A_1\neq 0$.
For the first possibility, we have the following MCs
\begin{equation}
\xi^0=\xi^0(x^a),~~~\xi^1=0,~~~\xi^2=c_1z+c_2,~~~\xi^3=-c_1y+c_3.
\end{equation}
When $A_1\neq 0$, we obtain
\begin{equation}
\frac{A_1(y,z)_{,kk}}{A_1(y,z)}=(\frac{T'_2}{2T_2\sqrt{T_1}})'
\frac{T_2}{\sqrt{T_1}}=\alpha,
\end{equation}
where $\alpha$ is an arbitrary constant which may be zero or
non-zero. The possibility $\alpha=0$ implies that
$\frac{T'_2}{T_2\sqrt{T_1}}=\beta$, an arbitrary constant and we
have the following MCs
\begin{eqnarray}
\xi^0&=&\xi^0(x^a),\\
\xi^1&=&\frac{1}{\sqrt{T_1}}(c_1y+c_2)z+c_3y+c_4),\\
\xi^2&=&-[(\frac{\beta}{4}y^2+\int{{\sqrt{T_1}}{T_2}dx})c_1z
+\frac{\beta c_2}{2}yz\nonumber\\
&+&(\frac{\beta}{4}(y^2-z^2)
+\int{{\sqrt{T_1}}{T_2}dx})c_3+\frac{\beta c_4}{2}y-c_5z+c_7,\\
\xi^3&=&-[(\frac{\beta}{4}z^2+\int{{\sqrt{T_1}}{T_2}dx})c_1y
+\frac{\beta c_3}{2}yz\nonumber\\
&+&(\frac{\beta}{4}(z^2-y^2)
+\int{{\sqrt{T_1}}{T_2}dx})c_2+\frac{\beta c_4}{2}z+c_5y+c_6.
\end{eqnarray}
For $\alpha\neq 0$, the MCs are given by
\begin{equation}
\xi^0=\xi^0(x^a),~~~\xi^1=0,~~~ \xi^2=c_0z+c_1,~~~
\xi^3=-c_0y+c_2.
\end{equation}
In the case 3(b), when $T_0=\gamma$ and $T_2=\delta$, where
$\gamma$ and $\delta$ are arbitrary constants, we have the
following MCs
\begin{eqnarray}
\xi^0&=&c_4y+c_5z+c_0,\\
\xi^1&=&\xi^1(x^a),\\
\xi^2&=&c_1z-\frac{\delta c_4}{\gamma}t+c_3,\\
\xi^3&=&-c_1y-\frac{\delta c_5}{\gamma}t+c_3.
\end{eqnarray}
If $T'_0\neq 0$ and $T'_2=0$, the MCs are given by
\begin{equation}
\xi^0=f(t),~~~\xi^1=-\frac{2T_0}{T'_0}\dot{f}(t),~~~
\xi^2=c_1z+c_2,~~~\xi^3=-c_1y+c_3.
\end{equation}
When $T'_2\neq 0$ and $\frac{T'_0T_2}{T_0T'_2}=\epsilon=0$, we
obtain the following MCs
\begin{eqnarray}
\xi^0&=&c_0,\\
\xi^1&=&-\frac{2T_2}{T'_2}(f'(u)+g'(v)),\\
\xi^2&=&f(u)+g(v),\\
\xi^3&=&-\iota(f(u)-g(v))+c_1,
\end{eqnarray}
where $u=y+\iota z$ and $v=y-\iota z$. For $T'_2\neq
0,~\epsilon\neq 0$ and $(\frac{T_0}{T_2})'\neq 0$, the proper MCs
are given by
\begin{eqnarray}
\xi_{(5)}&=&t\partial_t-\frac{2T_0}{T'_0}\partial_x,\nonumber\\
\xi_{(6)}&=&\iota(z\partial_y-y\partial_z).
\end{eqnarray}
This gives two proper MCs. If $T'_2\neq 0,~T_0=\lambda T_2$ and
$T'_0\neq 0$, we get
\begin{eqnarray}
\xi_{(5)}&=&\frac{1}{\lambda}[ty\partial_t-\frac{2T_2}{T'_2}y
\partial_x+\frac{1}{2}(y^2-z^2-\lambda t^2)\partial y
+yz\partial z],\nonumber\\
\xi_{(6)}&=&tz\partial_t-\frac{2T_2}{T'_2}z
\partial_x+yz\partial y-\frac{1}{2}(y^2-z^2
-\lambda t^2)\partial z,\nonumber\\
\xi_{(7)}&=&t\partial t-\frac{2T_2}{T'_2}\partial_x+y\partial y
+z\partial z,\nonumber\\
\xi_{(8)}&=&\frac{1}{2\lambda}(y^2+z^2-\lambda
t^2)\partial_t+\frac{2T_2}{T'_2}t \partial_x-ty\partial y
-tz\partial z,\nonumber\\
\xi_{(9)}&=&\frac{1}{\lambda}y\partial t-t\partial y,\nonumber\\
\xi_{(10)}&=&\frac{1}{\lambda}z\partial t-t\partial t.
\end{eqnarray}
which yields six proper MCs. Finally, when $T'_2\neq 0$ and
$(\frac{T'_0T_2}{T_0T'_2})'\neq 0$, we have four independent MCs
which are exactly the isometries of the plane symmetry. It is
interesting to note that the last three subcases of this case give
finite dimensional MCs even for the degenerate case.

\section{Examples Admitting Proper MCs}

In this section we construct examples which admit proper MCs for
the non-degenerate energy-momentum tensor. It can be seen from
Eq.(A3) that energy-momentum tensor will be non-zero when neither
of the metric functions $\nu$ and $\mu$ are constants. If we
choose these $\nu$ and $\mu$ such that $\nu=a\ln x+b,~~\mu=c\ln
x+d$, where a,b,c,d are constants such that $a\neq c$ then
\begin{equation}
\frac{T'_2}{T_2\sqrt{T_1}}=\frac{2(c-2)}{\sqrt{c(c+2a)}}
=constant=\alpha\neq 0,
\end{equation}
\begin{equation}
\frac{T'_0}{T_0\sqrt{T_1}}=\frac{2(a-2)}{\sqrt{c(c+2a)}}
=constant=\beta\neq 0.
\end{equation}
This shows that $\alpha\neq \beta$ as $a\neq c$ and hence the
metric
\begin{equation}
ds^2=x^adt^2-dx^2-x^c(dy^2+dz^2).
\end{equation}
admits five MCs.

If we choose $c=2$ and $a\geq 0$ but $a\neq 2$ in Eq.(82), it
admits six MCs. If we choose $a=2$ and $c\geq 0$ but $c\neq 2$ in
the metric given by Eq.(82), it yields seven MCs. Finally, if we
choose $\nu=ax=\mu$, then the constraint equations corresponding
to ten MCs are satisfied and we obtain the following metric
\begin{equation}
ds^2=e^{ax}dt^2-dx^2-e^{ax}(dy^2+dz^2).
\end{equation}
This is the well known anti-de Sitter metric.

\section{Discussion and Conclusion}

In a recent paper [12], some interesting results have been
obtained when we classify static spherically symmetric spacetimes
according to their energy-momentum tensor. In this paper, we have
extended the same procedure to classify static plane symmetric
spacetimes according to their MCs.

In the non-degenerate case, we obtain either four, five, six,
seven or ten independent MCs. These contain the usual four
isometries of the plane symmetry and the rest are the proper MCs.
For the degenerate energy-momentum tensor, most of the cases give
infinite dimensional MCs. The worth mentioning cases are those
where we have got finite number of MCs even when the
energy-momentum tensor is zero. We obtain three such different
cases having either four, six or ten independent MCs. The results
are
summarized in the form of tables given below. \\

\vspace{0.2cm}

{\bf {\small Table 1.}} {\small MCs for the Non-degenerate Case}

\vspace{0.1cm}

\begin{center}
\begin{tabular}{|l|l|l|}
\hline {\bf Cases} & {\bf MCs} & {\bf Constraints}
\\ \hline 1a & $6$ & $(\frac{T'_2}{T_2\sqrt{T_1}})'\neq 0,~
(\frac{T_0}{T_2})'=0$
\\ \hline 1b & $4$ &
$(\frac{T'_2}{T_2\sqrt{T_1}})'\neq 0,~(\frac{T_0}{T_2})'\neq 0$
\\ \hline 2ai* & $5$ &
$(\frac{T'_2}{T_2\sqrt{T_1}})'=0,~\frac{T'_2}{T_2\sqrt{T_1}}\neq
0,~(\frac{T_2}{T_0})'\neq 0,~\frac{T'_0}{T_0\sqrt{T_1}}\neq 0$
\\ \hline 2ai** & $7$ &
$(\frac{T'_2}{T_2\sqrt{T_1}})'=0,~\frac{T'_2}{T_2\sqrt{T_1}}\neq
0,~(\frac{T_2}{T_0})'\neq 0,~\frac{T'_0}{T_0\sqrt{T_1}}=0$
\\ \hline 2aii & $10$ &
$(\frac{T'_2}{T_2\sqrt{T_1}})'=0,~\frac{T'_2}{T_2\sqrt{T_1}}\neq
0,~(\frac{T_2}{T_0})'=0$
\\ \hline 2bi* & $10$ &
$(\frac{T'_2}{T_2\sqrt{T_1}})'=0,~\frac{T'_2}{T_2\sqrt{T_1}}=0,~
(\frac{(\sqrt{T_0})'}{T_1})'=0,~\frac{(\sqrt{T_0})'}{T_1}\neq 0$
\\ \hline 2bi** & $10$ &
$(\frac{T'_2}{T_2\sqrt{T_1}})'=0,~\frac{T'_2}{T_2\sqrt{T_1}}=0,~
(\frac{(\sqrt{T_0})'}{T_1})'=0,~\frac{(\sqrt{T_0})'}{T_1}=0$
\\ \hline 2bii*+ & $6$ & $
\begin{array}{c}
(\frac{T'_2}{T_2\sqrt{T_1}})'=0,~\frac{T'_2}{T_2\sqrt{T_1}}=0,~
(\frac{(\sqrt{T_0})'}{T_1})'\neq 0,\\
(\frac{T_0}{2}\sqrt{T_1}(\frac{T'_0}{T_0\sqrt{T_1}})')'=0,~
\frac{T_0}{2}\sqrt{T_1}(\frac{T'_0}{T_0\sqrt{T_1}})'=0
\end{array}
$\\ \hline 2bii*++ & $6$ & $
\begin{array}{c}
(\frac{T'_2}{T_2\sqrt{T_1}})'=0,~\frac{T'_2}{T_2\sqrt{T_1}}=0,~
(\frac{(\sqrt{T_0})'}{T_1})'\neq 0,\\
(\frac{T_0}{2}\sqrt{T_1}(\frac{T'_0}{T_0\sqrt{T_1}})')'=0,~
\frac{T_0}{2}\sqrt{T_1}(\frac{T'_0}{T_0\sqrt{T_1}})'\neq 0
\end{array}
$\\ \hline 2bii** & $4$ & $
\begin{array}{c}
(\frac{T'_2}{T_2\sqrt{T_1}})'=0,~\frac{T'_2}{T_2\sqrt{T_1}}=0,~
(\frac{(\sqrt{T_0})'}{T_1})'\neq 0,\\
(\frac{T_0}{2}\sqrt{T_1}(\frac{T'_0}{T_0\sqrt{T_1}})')'\neq 0
\end{array}
$\\ \hline
\end{tabular}
\end{center}

\vspace{0.2cm}

{\bf {\small Table 2}. }{\small MCs for the Degenerate Case (only
finite cases)}

\vspace{0.1cm}

\begin{center}
\begin{tabular}{|l|l|l|}
\hline {\bf Cases} & {\bf MCs} & {\bf Constraints}
\\ \hline 3bi & $6$ & $T_1=0,~T_j\neq 0(j=0,2,3),~T'_2\neq 0,~
\frac{T'_0T_2}{T_0T'_2}\neq 0,~(\frac{T_0}{T_2})'\neq 0$
\\ \hline 3bii & $10$ &
$T_1=0,~T_j\neq 0,~T'_2\neq 0,~T_0=\lambda T_2,~T'_0\neq 0$
\\ \hline 3biii & $4$ &
$T_1=0,~T_j\neq 0,~T'_2\neq 0,~\frac{T'_0T_2}{T_0T'_2}\neq 0$
\\ \hline
\end{tabular}
\end{center}

\vspace{0.2cm}

From these tables, it follows that each case has different
constraints on the energy-momentum tensor. Finally, we have
constructed some examples satisfying the given constraints.

When the rank of $T_a$ is 3, i.e. $T_1=0$, we obtain the following
metric
\begin{equation}
ds^2=e^\nu dt^2-dx^2-e^{-2\nu}(dy^2+dz^2),
\end{equation}
where $\nu$ is an arbitrary function of $x$ only. It can be easily
verified that this class of metrics represent perfect fluid dust
solutions. The energy-density for the above metrics is given as
\begin{equation}
\rho=(2\nu''-3\nu'^2)e^{\frac{\nu}{2}},
\end{equation}
It would be interesting to solve the constraints involved or more
examples should be constructed to check the dimensions of the MCs.

\newpage
\renewcommand{\theequation}{A\arabic{equation}}
\setcounter{equation}{0}
\section*{Appendix A}

The surviving components of the Ricci tensor are
\begin{eqnarray}
R_0\equiv
R_{00}&=&\frac{1}{4}e^\nu(2\nu''+\nu'^2+2\nu'\mu'),\nonumber \\
R_1\equiv R_{11}&=&-\frac{1}{4}(2\nu''+\nu'^2+4\mu''+2\mu'^2),\nonumber \\
R_2\equiv R_{22}&=&-\frac{1}{4}e^\mu(2\mu''+2\mu'^2+\nu'\mu'),\nonumber \\
R_{33}&=&R_{22}.
\end{eqnarray}
The Ricci scalar is given by
\begin{eqnarray}
R&=&\frac{1}{2}(2\nu''+\nu'^2+2\nu'\mu'+3\mu'^2+4\mu'').
\end{eqnarray}
Using Einstein field equations, the non-vanishing components of
energy-momentum tensor $ T_{ab} $ are
\begin{eqnarray}
T_0\equiv T_{00}&=&-\frac{1}{4}e^\nu(4\mu''+3\mu'^2),\nonumber \\
T_1\equiv T_{11}&=&\frac{1}{4}(\mu'^2+2\nu'\mu'),\nonumber \\
T_2\equiv T_{22}&=&\frac{1}{4}e^{\mu}(2\nu''+\nu'^2+\nu'\mu'
+\mu'^2+2\mu''),\nonumber \\
T_{33}&=&T_{22}.
\end{eqnarray}

\renewcommand{\theequation}{B\arabic{equation}}
\setcounter{equation}{0}
\section*{Appendix B}

The four independent KVs associated with the plane symmetric
spacetimes are given by [3]
\begin{eqnarray}
\xi_{(1)}=\partial_t,~~\xi_{(2)}=\partial_y,~~
\xi_{(3)}=\partial_z,~~ \xi_{(4)}=z\partial_y-y\partial_z.
\end{eqnarray}

\newpage

\begin{description}
\item  {\bf Acknowledgment}
\end{description}

I would like to thank Ministry of Science and Technology (MOST),
Pakistan for providing postdoctoral fellowship at University of
Aberdeen, UK.

\vspace{2cm}

{\bf \large References}

\begin{description}

\item{[1]} Katzin, G.H., Levine J. and Davis, W.R.: J. Math. Phys.
{\bf 10}(1969)617.

\item{[2]} Petrov, A.Z.: {\it Einstein Spaces} (Pergamon, Oxford
University Press, 1969).

\item{[3]} Stephani, H., Kramer, D., MacCallum, M.A.H.,
Hoenselaers, C. and Hearlt, E.: {\it Exact Solutions of Einstein's
Field Equations} (Cambridge University Press, 2003).

\item{[4]} Rcheulishrili, G.: J. Math. Phys. {\bf 33}(1992)1103.

\item{[5]} Coley, A.A. and Tupper, O.J.: J. Math. Phys. {\bf
30}(1989)2616.

\item{[6]} Hall, G.S., Roy, I. and Vaz, L.R.: Gen. Rel and Grav.
{\bf 28}(1996)299.

\item{[7]} Camc{\i}, U. and Barnes, A.: Class. Quant. Grav. {\bf
19}(2002)393.

\item{[8]} Carot, J. and da Costa, J.: {\it Procs. of the 6th
Canadian Conf. on General Relativity and Relativistic
Astrophysics}, Fields Inst. Commun. 15, Amer. Math. Soc. WC
Providence, RI(1997)179;\\
Carot, J., da Costa, J. and Vaz, E.G.L.R.: J. Math. Phys. {\bf
35}(1994)4832.

\item{[9]} Tsamparlis, M., and Apostolopoulos, P.S.: J. Math.
Phys. {\bf 41}(2000)7543.

\item{[10]} Sharif, M.: Nuovo Cimento {\bf B116}(2001)673;\\
Astrophys. Space Sci. {\bf 278}(2001)447.

\item{[11]} Camc{\i}, U. and Sharif, M.: Gen Rel. and Grav. {\bf
35}(2003)97;\\ Class. Quant. Grav. {\bf 20}(2003)2169.

\item{[12]} Sharif, M. and Sehar Aziz: Gen Rel. and Grav. {\bf
35}(2003)1091.

\item{[13]} Hall, G.S.: Gen. Rel and Grav. {\bf 30}(1998)1099.

\end{description}

\end{document}